\documentclass{ptapap}
\usepackage{color}

\usepackage{soul}

\usepackage{amsmath}
\usepackage{amssymb}
\usepackage{natbib}
\usepackage{float}
\usepackage{graphicx}
\usepackage{subcaption}
\usepackage{multicol}
\usepackage{caption}

\author{Aleksandra Kotek}[CAMK]
\author{Miljenko \v{C}emelji\'{c}}[CAMK,AS]
\author{W\l odek Klu\'{z}niak}[CAMK]
\author{Deepika A. Bollimpalli}[CAMK,USA]
\affil[CAMK]{Nicolaus Copernicus Astronomical Center, Polish Academy of Sciences, Bartycka 18, 00--716 Warsaw, Poland}
\affil[UJ]{Astronomical Observatory, Jagiellonian University, ul. Orla 171, 30-244 Kraków, Poland}
\affil[AS]{Academia Sinica, Institute of Astronomy and Astrophysics, P.O. Box 23-141, Taipei 106, Taiwan}
\affil[USA]{Department of Physics and Astronomy, College of Charleston, Charleston, SC 29424,  USA}
\title{Asymmetric Jet launching}

\begin{document}

\maketitle

\begin{abstract}
In resistive and viscous magnetohydrodynamical (MHD) simulations we obtain axial jets launched from the innermost magnetosphere of a star-disk system. We found that in a part of the parameter space continuous asymmetric jets, which are propagating in opposite directions, are launched. We compare the speed of propagation and rotation of obtained jets with recent observational results. 
\end{abstract}

\section{Introduction}
\begin{figure}
\centering 
\includegraphics[width=0.65\columnwidth]{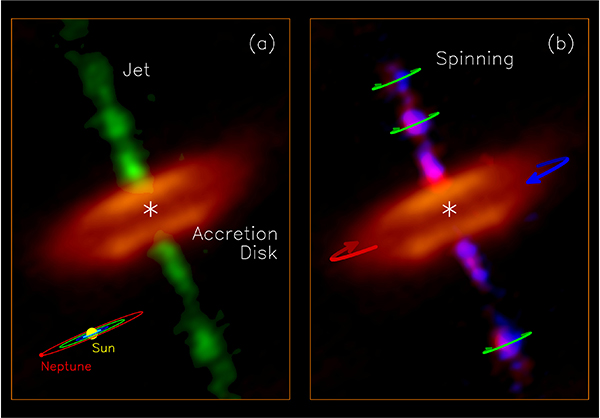}
\includegraphics[width=0.32\columnwidth]{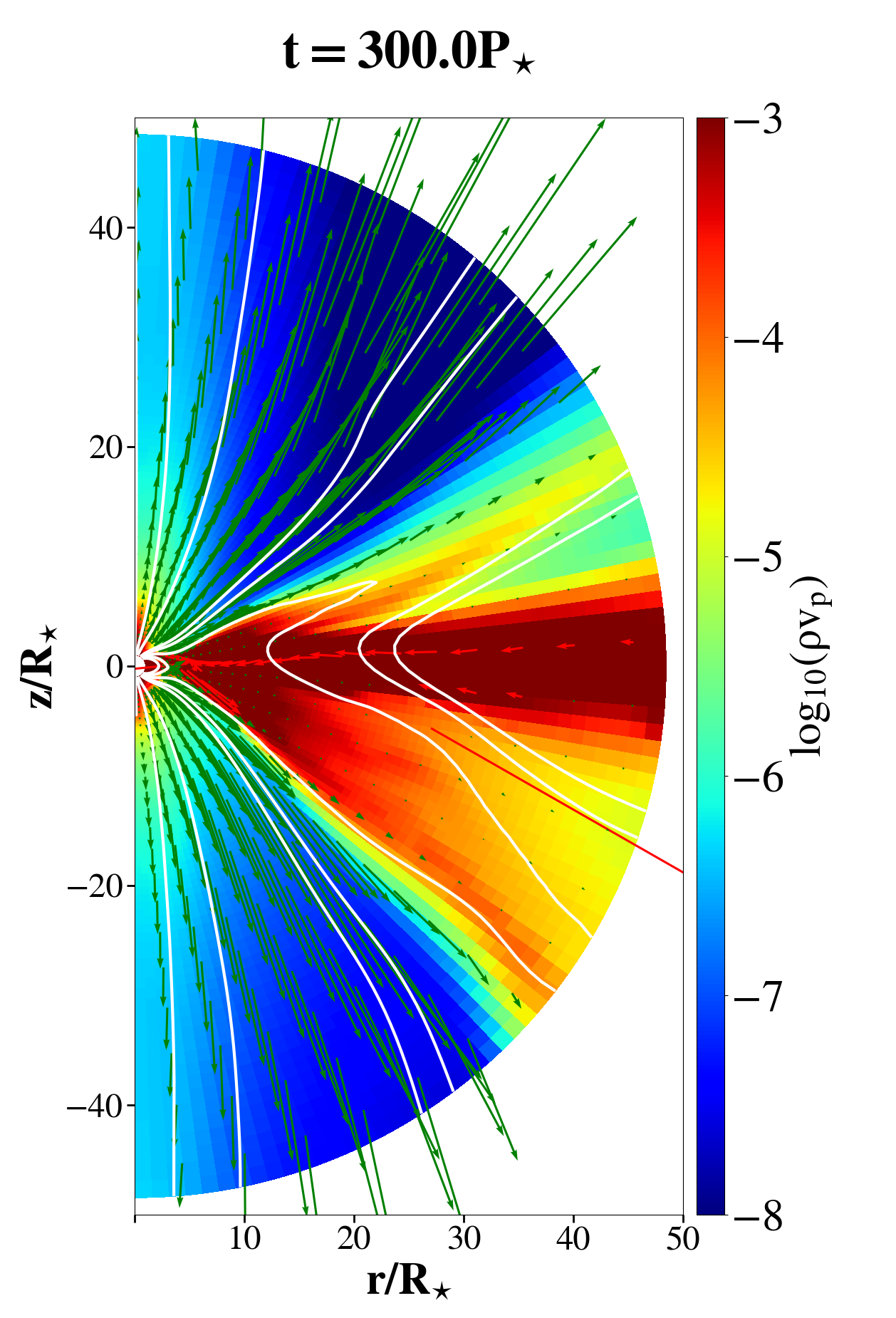}
\caption{Left panels: The measurement of HH 212 protostellar system implies a jet launching based on the magneto-centrifugal mechanism, which connects the properties of the jet measured at large distances with those at its base through energy and angular momentum conservation.
Right panel: Matter flux in a logarithmic color grading. Normalization of the velocity vectors in the disk (red) is hundred times the velocity in the magnetosphere above the disk (green), as the velocity in the disk is much smaller than the velocity in the corona. With white solid lines is shown a sample of the magnetic field lines.}
\label{fig:f1}
\end{figure}

We performed a parameter study for the slowly rotating Young Stellar Objects
(YSOs), to find the cases when axial jets are launched from the
star-disk system magnetosphere. Asymmetric jets are launched in the
opposite directions above the stellar surface, with different propagation
and rotation speeds and different matter fluxes. Results of
MHD simulations (\cite{R09}, \citet{ZF13},
\cite{cem19}) are helpful in explaining the launching mechanism of jets and outflows. Direct observations  \cite[e.g.][]{cflee17} are also becoming available -- see Fig.~\ref{fig:f1} left panels.

\section{Numerical setup}
Our setup is an extension of Čemeljić (2019). The disk set up is following \cite{KK00} solution, with the addition of hydrostatic, initially non-rotating corona above the rotating star. The viscosity and resistivity are parameterized by the \cite{ShaSun} prescription, with a dipole stellar field. We use constrained transport method, together with split-field approach, in which only changes from the initial stellar magnetic field are evolved in time. 

\section{Asymmetric jets}
We perform 2D-axisymmetric star-disk simulations in a complete [0,$\pi$] meridional half-plane, in the resolution $R\times\theta = [125\times100]$ grid cells, reaching the maximal radius of 50 stellar radii--see Fig.~\ref{fig:f1} right panel. The PLUTO code \citep{m07} with a logarithmic stretched radial grid and uniform latitudinal grid in spherical coordinates is used. We obtain quasi-stationary solution in our simulation with asymmetric axial jets launched in opposite directions from the magnetosphere of a star-disk system. Such results can be compared to the observations.

\begin{figure}
\centering 
\includegraphics[height=0.4\columnwidth,width=0.48\columnwidth]{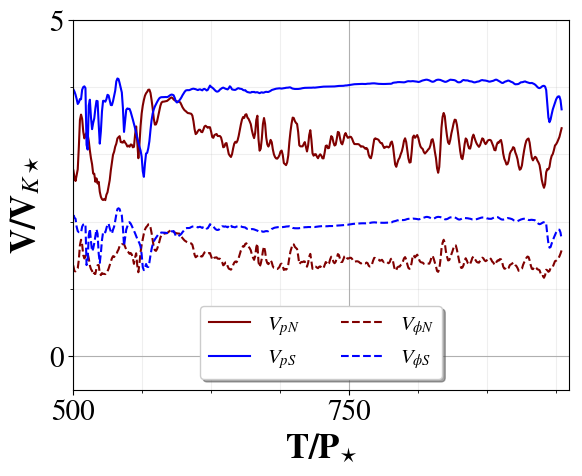}
\includegraphics[height=0.4\columnwidth,width=0.48\columnwidth]{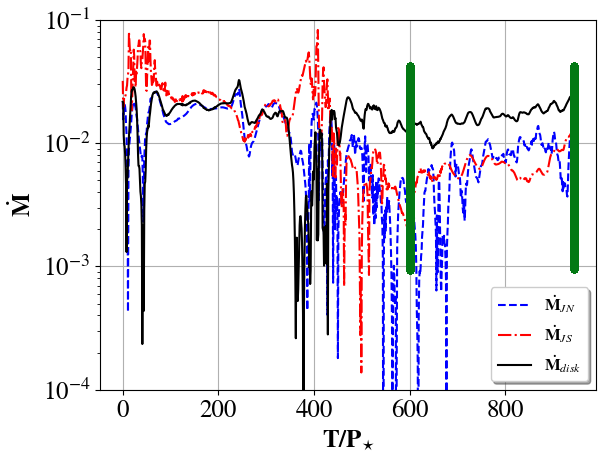}
\caption{
Left panel: Distribution of poloidal and azimuthal components of the jet velocities in the units of Keplerian velocity at the stellar equator.
Velocities of the Northern jet are shown with blue, and of the Southern jet with red lines.
Right panel: Mass flux along the line at 
$R=25 R_{*}$ in jets and disk, during the whole simulation, in the units of $10^{-7} M_\odot$ yr$^{-1}$. With the black solid line the mass flux through the disk
is shown, and with blue and red dashed lines through the Northern and Southern jets, respectively.} 

    \label{fig:vel}
\end{figure}

The distribution of jet velocities during the quasi-stationary state in our simulation is shown in Fig.~\ref{fig:vel} left panel. Time is measured in the number of stellar rotations. Two upper lines show the jet propagation speed, while the two bottom lines show the jet rotational velocity. 

After relaxation from the initial conditions, a quasi-stationary state is reached, as shown in Fig.~\ref{fig:vel} right panel, marked with the thick green vertical solid lines. This figure presents mass flux through the Northern and Southern jets, which is different for both cases. 
In both jets, it is of the order of few per cents of the disk accretion rate. Time again is measured in the number of stellar rotations. Velocities from interval 600 to 950 $TP_*^{-1}$ are averaged for comparisons in the parameter study. 

Below we present the table with the parameter space for the slowly rotating 
YSOs with different magnetic fields strengths, disk anomalous resistivity coefficients 
$\alpha_{\rm m}$ and rotational velocities $\Omega_{\star} / \Omega_{\rm br}$. 
In all the cases the anomalous viscosity parameter is $\alpha_{\rm v}=1$.  The annotations are as
follows: $x$ - simulation still to be performed, $Y$ - asymmetric jets are
present, and $N$ - no jets.
\begin{multicols}{2}

{
$\begin{array}{ | c | c | c | c | c |}     
\hline
\alpha_{\rm m}= & 0.1 & 0.4 & 0.7 & 1 \\
\hline\hline
\multicolumn{5}{c}{B_\star=250~{\rm G}} \\
\hline
 \Omega_\star/\Omega_{\rm br} & & & & \\
\hline

0.5 & Y & Y & Y & Y \\ 
0.8 & x  & Y  & Y  & Y  \\

\hline\hline     
\multicolumn{5}{c}{B_\star=500~{\rm G}} \\
0.2 & x & x & Y  & Y  \\
0.5 & Y & Y & N & Y \\ 
0.8 & Y & Y & Y & Y \\
 \hline\hline 
\end{array} $}

\columnbreak 
{
$\begin{array}{ | c | c | c | c | c |}     
\hline
\alpha_{\rm m}= & 0.1 & 0.4 & 0.7 & 1 \\
\hline\hline
\multicolumn{5}{c}{B_\star=750~{\rm G}} \\
\hline
 \Omega_\star/\Omega_{\rm br} & & & & \\
\hline
    
0.5 & Y & N & N & Y \\ 
0.8 & N & Y & N & Y \\

\hline\hline    
\multicolumn{5}{c}{B_\star=1000~{\rm G}} \\
0.2 & x & x & x & x \\
0.5 & N & x & N & N \\ 
0.8 & Y & Y & N & N \\  
 \hline\hline  
\end{array} $
}

\end{multicols}

\section{Conclusions}
In our numerical simulations, we obtained asymmetric jets launched from the magnetosphere of a star-disk system.
We have shown the preliminary results of a parameter study for  which the 
parameter space is determined in our viscous and resistive MHD simulations in which
axial jets are launched. Our results can be directly compared with observations.
\acknowledgements{We thank A. Mignone and his team of contributors for the possibility to use the PLUTO code, and ASIAA/TIARA and CAMK PAN for use of their Linux clusters XL and CHUCK, respectively. Work in Warsaw is funded by the Polish NCN grant No. 2013/08/A/ST9/00795. MČ developed the PLUTO setup under ANR Toupies funding in CEA Saclay, France, and also acknowledges Croatian HRZZ grant IP-2014-09-8656.}

\bibliographystyle{ptapap}
\bibliography{olaPTA2019_corr}

\begin{thebibliography}{7}
\providecommand{\natexlab}[1]{#1}
\providecommand{\url}[1]{\texttt{#1}}
\providecommand{\urlprefix}{URL }
\providecommand{\eprint}[2][]{\url{#2}}

\bibitem[{{Kluzniak} \& {Kita}(2000)}]{KK00}
{Kluzniak}, W., {Kita}, D., \emph{arXiv e-prints} astro-ph/0006266 (2000)

\bibitem[{{Lee} et~al.(2017)}]{cflee17}
{Lee}, C.-F., et~al., \emph{Nature Astronomy} \textbf{1}, 0152 (2017)

\bibitem[{{Mignone} et~al.(2007)}]{m07}
{Mignone}, A., et~al., \emph{\apjs} \textbf{170}, 228 (2007)

\bibitem[{{Romanova} et~al.(2009){Romanova}, {Ustyugova}, {Koldoba}, \&
  {Lovelace}}]{R09}
{Romanova}, M.~M., {Ustyugova}, G.~V., {Koldoba}, A.~V., {Lovelace}, R.~V.~E.,
  \emph{\mnras} \textbf{399}, 4, 1802 (2009)

\bibitem[{{Shakura} \& {Sunyaev}(1973)}]{ShaSun}
{Shakura}, N.~I., {Sunyaev}, R.~A., \emph{\aap} \textbf{500}, 33 (1973)

\bibitem[{{{\v{C}}emelji{\'c}}(2019)}]{cem19}
{{\v{C}}emelji{\'c}}, M., \emph{\aap} \textbf{624}, A31 (2019)

\bibitem[{{Zanni} \& {Ferreira}(2013)}]{ZF13}
{Zanni}, C., {Ferreira}, J., \emph{\aap} \textbf{550}, A99 (2013)

\end{thebibliography}

\end{document}